\documentclass{article}
\def\be{\begin{equation}}
\def\ee{\end{equation}}

\def\bea{\begin{eqnarray}}
\def\eea{\end{eqnarray}}

\def\del{\delta}

\newcommand{\dif}{{\mathrm{d}}}

\def\d{\partial}

\newcommand{\refeq}[1]{\mbox{(\ref{#1})}}
\newcommand{\ltsim}{\lower3pt\hbox{$\, \buildrel < \over \sim \, $}}
\newcommand{\gtsim}{\lower3pt\hbox{$\, \buildrel > \over \sim \, $}}
\usepackage{graphics,epsfig}
\begin{document}
\begin{flushright}
hep-ph/0310352 \\
UG-FT-159/03, CAFPE-29/03\\
October 2003\\
\end{flushright}
\vspace*{5mm}
\begin{center}

{\Large{\bf Bulk fields with brane terms}
\footnotemark[1]\footnotetext[1]{Presented at 
HEP2003, International Europhysics Conference on High
Energy Physics (July 17-23, 2003) Aachen, Germany.}}

\vspace{1.4cm}
{\sc F. del Aguila$^1$, M. P\'erez-Victoria$^2$ and J. Santiago$^3$}\\
\vspace{.5cm}
{\it $^1$Centro Andaluz de
F\'\i sica de Part\'\i culas Elementales (CAFPE)}\\
{\it and Departamento de F\'\i sica Te\'orica y del Cosmos}\\
{\it Universidad de Granada, E-18071 Granada, Spain}\\
\vspace{.5cm}
{\it $^2$Dipartimento di Fisica ``G. Galilei'', Universit\`a di
  Padova and}\\
{\it INFN, Sezione di Padova, Via Marzolo 8, I-35131 Padua, Italy}\\
\vspace{.5cm}
{\it $^3$IPPP, Centre for Particle Theory}\\
{\it University of Durham}\\
{\it DH1 3LE, Durham, U.K.}\\

\end{center}
\vspace{1.cm}
\begin{abstract}

In theories with branes, bulk fields get in general divergent
corrections localized on these defects. Hence, the corresponding brane
terms are renormalized and should be included in the effective
theory from the very beginning. We review the phenomenology associated
to brane kinetic terms for different spins and backgrounds,
and point out that renormalization is required already at
the classical level.

\end{abstract}
\section{Introduction}
\label{intro}
Models with extra dimensions and branes allow for new ways of
addressing longstanding questions within the Standard Model (SM), such
as the hierarchy problem or the structure of fermion
masses and mixings. 
(See \cite{Hewett} for a review of their phenomenology.) 
These models are
nonrenormalizable and must be understood as effective theories valid
at energies below a certain cutoff $\Lambda$. One can distinguish
fields which propagate in all dimensions (``bulk fields'') from those
which propagate only on the branes. Both kinds of fields can
couple via operators which are necessarily localized on the
branes~\cite{Horava96,Mirabelli98}. One can also consider brane
localized operators involving only 
bulk fields. Since Poincar\'e invariance is broken by the presence of
the brane (through brane fields or an orbifold projection), these
operators typically receive divergent radiative corrections which
require renormalization~\cite{Dvali00,Georgi01}. Thus, their
coefficients run and cannot be set to 
zero at all scales. It is therefore natural to include the brane terms
from the very beginning in the (tree-level) effective action.
In particular, one should expect the presence of Brane
localized Kinetic Terms for all fields in the bulk (BKT).  
Their impact on phenomenology has been studied recently for different
fields and backgrounds in a number of papers. It is our
purpose here to review some of the main results in these
works. We will also emphasize the breakdown of the low-energy
expansion brought about by certain BKT, and comment on how
to make sense of the theory.
\section{Phenomenology of Brane Kinetic Terms}
\label{sec:1}
The action for a bulk scalar in 5D with BKT has the form
\begin{equation}
S = \int \dif^4 x \int \dif y 
\big(\partial_M \phi^\dagger \partial^M \phi + 
a^\phi_I \delta_I \partial_\mu \phi^\dagger \partial^\mu \phi
 + \dots \big) , 
\label{scalaraction}
\end{equation}
where $\delta_I\equiv \delta(y-y_I)$, $y_I$ are the location of the
branes and a sum over $I$ is implicit. In the $1/\Lambda$ expansion,
the first term is order zero, whereas the second one is order
$1/\Lambda$. In order for this expansion to be well defined,  
in the limit $a_I^\phi \rightarrow 0$ the observable predictions should
approach the ones for $a_I^\phi=0$. However, this is not true for all
BKT~\cite{Aguila03a}. We will comment on ``singular'' BKT
later in Section~\ref{sec:5} and stick for the moment to BKT which behave
smoothly when their coefficients go to zero. 
The phenomenological implications of these terms depend on the type of
field, but they look pretty similar for different backgrounds (in
compact scenarios). We discuss gauge fields, gravitons, and fermions
in turn. We consider the coefficients of the BKT as free parameters of
arbitrary size. Observe that even if they are small
they can significantly modify the physics by breaking
degeneracies among KK modes~\cite{Cheng02}.
%
\subsection{Gauge fields}
\label{sec:2}
BKT for gauge bosons in infinite extra dimensions have been
considered in~\cite{Dvali00b}, with features analogous to the ones
described for the graviton below. In this subsection we consider only
bulk gauge 
bosons propagating in compact extra dimensions. The case of $M_4
\times S^1/Z_2$ has been studied in \cite{Carena02}. The corresponding 
lagrangian is
\begin{equation}
\mathcal{L} =  -\frac{1}{4}
\big(\mathrm{tr} F_{MN}F^{MN} + a^A_I\del_I 
\mathrm{tr} F_{\mu\nu}F^{\mu\nu} \big) + \dots ,  
\label{gaugeLag}
\end{equation}
with $I = 0, \pi$, and then the branes located at $y_I = IR$
with $R$ the orbifold radius.
For $a^A_0 + a^A_{\pi} \le -2\pi R$ there are ghosts, 
and for $-2\pi R < a^A_I < 0$ tachyons, so we must take $a^A_I \geq
0$. 
>From a 4D point of view, this theory is described by a 
Kaluza-Klein (KK) tower of spin 1 fields,
$A_{\mu} (x,y) = \sum_{n=0}^\infty \frac{f^A_n(y)}{\sqrt{2\pi R}}
A_\mu^{(n)}(x) $, 
with eigenfunctions $f^A_n(y)$ of eigenvalue 
$m_n$ diagonalizing \refeq{gaugeLag}. 
The BKT modify the diagonalization and normalization equations and, 
hence, the eigenfunctions and eigenvalues of the heavy modes. 
If only $a_0$, say, is nonvanishing, the latter decrease, 
whereas the former tend to decouple from the $y_0$ brane. 
Hence, they are easier to reach at large colliders 
but, if matter is located at $y_0$, their production cross sections
are smaller. See \cite{Carena02,Aguila03c} for the detailed dependence.
For equal non-zero BKT at both branes, the lightest KK boson 
acts as a collective mode which becomes light and
non decoupling for large $a^A_I$. 

The integration of the KK modes gives four-fermion operators 
whose strengths are constrained by precision electroweak (LEP) data 
\cite{Rizzo99}. The reduction of the KK masses and couplings in the presence
of BKT produces as a net result a relaxation of the corresponding
limits, allowing for a lower compactification scale $1/R$
\cite{Carena02,Aguila03c}. 

An analogous behaviour is shared by warped backgrounds. 
A detailed discussion of the implications of non-vanishing BKT 
for the Randall-Sundrum (RS) model \cite{Randall99} 
can be found in \cite{Davoudiasl03a,Carena03}, with similar 
conclusions. In particular, models with lower KK masses, maybe 
at the reach of next colliders, can be accommodated within 
experimental bounds. 

Finally, we should mention that BKT must be kept small in GUT
scenarios such as~\cite{Hall02}, as they can spoil gauge
unification. On the 
other hand, they can be helpful in improving some features of models
of electroweak symmetry breaking~\cite{Scrucca03}. 


\subsection{Graviton}
\label{sec:3}
Since gravity necessarily propagates in all the space, we must expect
graviton BKT to be a generic feature of models with branes. 
These terms were first introduced to allow for an effective 4D
Newton's law in 5D models with an infinite extra
dimension~\cite{Dvali00}.   
The addition to the 5D curvature term $R^{(5)}$
of a brane term  $a^g\del_0 R^{(4)}$ 
gives rise to Newton's law $V(r)\sim \frac{1}{r}$ at small distances
$r\ll a^g$, and to a quadratic power law $V(r)\sim \frac{1}{r^2}$ at
larger ones $r\gg a^g$. Then, on phenomenological grounds the
transition length $a^g$ must be of the size of the present
horizon. This modification of 4D gravity at large distances is very
interesting since it allows for solutions of the cosmological
constant problem, evading no-go theorems for theories which are 4D in
the infrared~\cite{Deffayet,Dvali02}. This model, however, 
presents strong quantum effects at distances 
$(\frac{a^{g 2}}{M_P})^{1/3} \sim 1000$ km, where 
$M_P$ is the Planck mass, requiring new physics at 
this scale and then making the proposed model 
incomplete \cite{Luty03,Rubakov03}. (According to~\cite{Gabadadze03},
however, this might be a calculational artifact.)

For a compactified extra dimension, the analysis of BKT is similar to
the one of gauge bosons discussed above. The RS model with BKT, 
\begin{equation}
\mathcal{L} = \frac{M_5^3}{4} \sqrt{-G} 
\big( R^{(5)} +  a^g_I \delta_I  R^{(4)} + \dots \big) , 
\end{equation}
has been analysed in \cite{Davoudiasl03b}. Again, the 4D KK gravitons
may be  
significantly ligther for large $a^g_I$ than for vanishing BKT, 
and their couplings to matter on the brane are smaller.
They may avoid detection at large colliders 
as their widths may be too narrow to be observable. 
They also give rise to four-fermion operators when 
integrated out, but in this case their strengths are independent 
of $a^g_I$ for a wide range of parameters.
Finally, gravitational BKT are also useful in model building. For
instance, they have been invoked in a model of gravity
mediated supersymmetry breaking to avoid negative squared masses for
scalars~\cite{Rattazzi03}.

\subsection{Fermions}
\label{sec:4}
The phenomenological consequences of BKT for bulk 
fermions in a 5D theory compactified on 
$M_4 \times S^1/Z_2$ were presented in \cite{Aguila03b}. 
The lagrangian is
\begin{equation}
\mathcal{L} = (1+ a_I^L \delta_I)
\bar{\Psi}_L \mathrm{i} \gamma^\mu D_\mu \Psi_L+\ldots .
\label{fermionLag}
\end{equation}
Again, the KK wave functions and masses are modified. As the new terms 
multiply the covariant derivative, the 4D 
gauge couplings for the different KK states get corrections both 
from the new wave functions and from the new 
effective couplings 
\begin{equation}
g^{(mnr)} = \frac{g_5}{\sqrt{2\pi R}}\int \mathrm{d}y\;
(1+ a^L_I  \delta_I )
\frac{ f_m^L f_n^L f_r^A}{2\pi R} ,
\label{gaugecouplings}
\end{equation}
where $L, A$ label fermion and gauge boson wave functions, 
respectively, and $g_5$ is the 5D gauge coupling. By gauge invariance,  
gauge bosons must propagate in the bulk if fermions do.
Hence, the integration of the heavy KK gauge bosons 
give new contributions to four-fermion operators bounded 
by electroweak precision (LEP) data \cite{Rizzo99}. 
Using \refeq{gaugecouplings} and the corresponding experimental 
limits one can estimate the exclusion region for the 
compactification scale as a function of $a^L_I$ 
\cite{Aguila03c,Aguila03b}.
The constraints become stronger with growing BKT, in 
contrast with the situation in Sections~\ref{sec:2} 
and \ref{sec:3}. 

The observed fermions are zero modes which get their masses  
through Yukawa couplings after the spontaneous symmetry 
breaking of the SM. These couplings also generate mixings 
with the heavy KK fermions. Hence, a precise determination of 
the charged current mixing matrix constrains the mixing and 
masses of these exotic fermions, and thus the compactification 
scale \cite{Aguila02}. In the presence of BKT their masses and mixings
can be reduced, with the net effect of relaxing the 
experimental constraints. As a consequence there is room for 
producing new vector-like fermions at future colliders 
in many models \cite{Exoticproduction}.
\section{General brane kinetic terms}
\label{sec:5}
The effects of the BKT in Section \ref{sec:1} are in general 
sizable only when the coefficients are large. Then, higher  
order terms in the effective theory may have to be taken into  
account. In general, one must also include BKT with derivatives
orthogonal to the brane. These have a
nonanalytical behaviour, in the sense that they give big effects for
arbitrarily small coefficients. A detailed discussion can be found in
\cite{Aguila03a}.
The most general kinetic terms for scalar, gauge bosons and
fermions read 
\begin{equation}
\begin{array}{c}
\mathcal{L}=
(1+ a^\phi_I  \delta_I) \partial_\mu
\phi^\dagger \partial^\mu \phi
- (1+c^\phi_I  \delta_I) \partial_y \phi^\dagger \partial_y \phi \\
+ \frac{ b_I^\phi  }{2} \delta_I (\phi^\dagger
  \partial_y^2 \phi +
  \partial_y^2   \phi^\dagger \phi) , \\
\end{array}
\label{scalarexample}
\end{equation}
\begin{equation}
\begin{array}{c}
\mathcal{L} = -\frac{1}{4} (1+ a_I^A \delta_I )
F_{\mu\nu}F^{\mu\nu} \\
 -\frac{1}{2} (1+ c^A_I \delta_I)
F_{5\nu}F^{5\nu} , \\
\end{array}
\end{equation}
and
\begin{equation}
\begin{array}{c}
\mathcal{L}= 
(1+  a^L_I \delta_I )
\bar{\Psi}_L \mathrm{i} \not \! \partial \Psi_L
+ (1+  a^R_I  \delta_I)
\bar{\Psi}_R \mathrm{i} \not \! \partial \Psi_R \\
-(1+ b_I^\psi \delta_I ) \bar{\Psi}_L \partial_y \Psi_R
- b_I^\psi  \delta_I \big( \partial_y \bar{\Psi}_R\big)
\Psi_L \\
+(1+ c_I^\psi  \delta_I ) \bar{\Psi}_R \partial_y
\Psi_L + c_I^\psi \delta_I 
\big( \partial_y \bar{\Psi}_L\big) \Psi_R , \\
\end{array}
\end{equation}
respectively. Let us discuss for illustration the 
scalar case in Eq. \refeq{scalarexample} 
with BKT at $y_0$ only.
For $b_0^\phi \neq 0$, the KK masses and wave functions diagonalizing 
the kinetic terms turn out to be independent of 
$a_0^\phi$, $b_0^\phi$ and $c_0^\phi$. Furthermore, their limits when 
$b_0^\phi \rightarrow 0$ do not coincide with their values at 
$b_0^\phi = 0$, which in particular have a non-trivial dependence on
$a_0^\phi$. This translates into dramatic changes in the
observables for arbitrarily small $b_0^\phi $. This singular behaviour 
comes from the fact that we are considering branes of zero width. In
perturbation theory, it manifests itself as singular 
products of delta functions at the origin. Taken at its faith value,
this is catastrophic, as it means that all models in which these
operators can be induced are badly defined as effective field
theories. Fortunately, the singular behaviour can be smoothed
down. The obvious way is to consider branes with finite
thickness~\cite{Kolanovic03a,Kolanovic03b}. However, predictions will
then depend on the particular 
profile of the brane. Another possibility is to go beyond
regularization of the brane, and renormalize the theory
at the classical level through the introduction of higher
order counterterms which cancel the
singularities~\cite{Aguila03a}. For instance, the second order
counterterm 
\begin{equation}
\begin{array}{c}
\mathcal{L}_{\mathrm{ct}}= \frac{ b_0^{\phi 2}}{4}
\big\{-\delta_0^2(y)  \d_\mu
\phi^\dagger \d^\mu \phi  
- \d_y \big[ \delta_0(y) 
  \phi^\dagger \big]
\d_y \big[ \delta_0(y)  \phi \big] \\
 + \delta_0^2(y)  \big(\phi^\dagger
\d_y^2 \phi + \d_y^2 \phi^\dagger \phi \big)\big\} , 
\end{array}
\end{equation}
makes the limit $b_0^\phi \rightarrow 0$ well-defined to second
order. At this order, the effect of the singular BKT can be absorbed
into a redefinition of the coefficients of the nonsingular BKT. 
It is plausible that, after
proper renormalization, the effects of all possible BKT can be
parametrized only by the coefficients of the nonsingular ones. If this
is the case, their phenomenology will reduce to the one discussed above.

To finish, let us mention that the situation in supersymmetric
theories is similar, except for the fact that some of the possible BKT
are stable against radiative corrections~\cite{Aguila03a} (see
also~\cite{Hebecker01,Kyae02}).

\end{document}